\documentclass{aa}                             
\usepackage{graphicx}
\usepackage{array}
\usepackage{lscape}                                
\usepackage{txfonts}
\usepackage{longtable}
\usepackage{color}
\usepackage{ulem}    
\usepackage{times}
\newcommand{\ioni}[2]{{#1\,\sc{#2}}}
\newcommand{\nioni}[2]{{[#1\,\sc{#2}]}}
\newcommand{\abun}[1]{{$\epsilon_{\rm Si}$}}

\def\msun{M$_{\odot}$}

\newcommand{\Ha}{H$\alpha$}
\newcommand{\Hb}{\ifmmode {\rm H}\beta \else H$\beta$\fi}

\newcommand{\hii}{H\,{\sc ii}}

\newcommand{\nii}{[N\,{\sc ii}]}
\newcommand{\Oi}{[O\,{\sc i}]\,$\lambda$6300}
\newcommand{\oi}{[O\,{\sc i}]}
\newcommand{\Oii}{[O\,{\sc ii}]\,$\lambda$3727}

\newcommand{\oii}{[O\,{\sc ii}]}
\newcommand{\Oiii}{[O\,{\sc iii}]\,$\lambda$5007}

\newcommand{\oiii}{[O\,{\sc iii}]}

\newcommand{\neiii}{[Ne\,{\sc iii}]}

\newcommand{\sii}{[S\,{\sc ii}]}

\newcommand{\siii}{[S\,{\sc iii}]}

\newcommand{\rOii}{[O\,{\sc ii}]\,$\lambda$3726/3729}

\newcommand{\rOiii}{[O\,{\sc iii}]\,$\lambda$4363/5007}
\newcommand{\rNii}{[N\,{\sc ii}]\,$\lambda$5755/6584}

\newcommand{\Hp}{H$^{+}$}

\newcommand{\Op}{O$^{+}$}
\newcommand{\Opp}{O$^{++}$}

\begin{document}
%
\title{No temperature fluctuations in the giant \hii\ region H\,1013 }

\author{G. Stasi\'nska\inst{1}, C. Morisset\inst{2}, S. Sim\'on-D\'iaz\inst{3,4}, F. Bresolin\inst{5}, D. Schaerer\inst{6,7}, B. Brandl\inst{8}}

\institute{LUTH, Observatoire de Paris, CNRS, Universit\'e Paris Diderot; Place Jules Janssen 92190 Meudon, France\\
\email{grazyna.stasinska@obspm.fr}
\and
Instituto de Astronom{\'\i}a, Universidad Nacional Aut\'onoma de M\'exico, Apdo. Postal 70264, M\'ex. D.F., 04510 M\'exico\\
\email{Chris.Morisset@Gmail.com}
\and
Instituto de Astrof\'isica de Canarias, E-38200 La Laguna, Tenerife, Spain            
             \and
             Departamento de Astrof\'isica, Universidad de La Laguna, E-38205 La Laguna, Tenerife, Spain\\
              \email{ssimon@iac.es}
\and
Institute for Astronomy, 2680 Woodlawn Drive, Honolulu, HI 96822, USA\\
\email{bresolin@ifa.hawaii.edu}
\and
Geneva Observatory, University of Geneva, 51, Ch. des Maillettes, CH-1290 Versoix, Switzerland
\and
CNRS, IRAP, 14 Avenue E. Belin, F-31400 Toulouse, France
\and
Leiden Observatory, Leiden University, P.O. Box 9513, NL-2300 RA Leiden, Netherlands
          }
		   
\offprints{grazyna.stasinska@obspm.fr}

\date{Submitted/Accepted}

\titlerunning{H\,1013}
\authorrunning{Stasi\'nska et al.}

%
\abstract{
While collisionally excited lines in \hii\ regions allow one to easily  probe the chemical composition of the interstellar medium in galaxies, the possible presence of important temperature fluctuations casts some doubt on the derived abundances.
To provide new insights into this question, we have carried out a detailed study of a giant \hii\ region, H\,1013, located in the galaxy M101, for which many observational data exist and which has been claimed to harbour temperature fluctuations at a level of $t^2 = 0.03-0.06$. 
We have first complemented the already available optical observational datasets with a mid-infrared spectrum obtained with the \textit{Spitzer Space Telescope}. Combined with optical data, this spectrum provides unprecedented information on the temperature structure of this giant \hii\ region. A preliminary analysis based on empirical temperature diagnostics suggests that temperature fluctuations should be quite weak. However, only a detailed photoionization analysis taking into account the geometry of the object and observing apertures can make a correct use of all the observational data. We have performed such a study using the Cloudy\_3D package based on the photoionization code Cloudy. We have been able to produce photoionization models constrained by the observed H$\beta$ surface brightness distribution and by the known properties of the ionizing stellar population than can account for most of the line ratios within their uncertainties. Since the observational constraints are both strong and numerous, this argues against the presence of significant temperature fluctuations in H\,1013.  The oxygen abundance of our best model  is 12 + log O/H = 8.57, as opposed to the values of 8.73 and 8.93 advocated by Esteban et al. (2009) and Bresolin (2007), respectively, based on the significant
temperature fluctuations they derived.

However, our model is not able to reproduce the intensities of the  oxygen recombination  lines  observed by Esteban et al. (2009), as well as  the very low Balmer jump temperature inferred by Bresolin (2007). We have argued that the latter might  be in error, due to observational difficulties. On the other hand, the discrepancy between model and observation as regards the recombination lines cannot be attributed to observational uncertainties and requires an explanation other than temperature fluctuations.
\keywords{ISM: abundances -- ISM: H{\sc ii} regions -- ISM: individual objects:
H\,1013 -- galaxies: individual: M 101}
}
%
\maketitle
%
%
\section{Introduction}\label{sec:intro}

Giant \hii\ regions are among the most effective  probes of the chemical composition of the present-day interstellar medium in galaxies. Methods to derive abundances in these objects have been devised many years ago (Aller \& Menzel 1945) and are straightforward. If the electron temperature can be measured using a temperature indicator such as the emission line ratio \rOiii\ or \rNii, the ionic abundances can be obtained directly from the intensities of lines emitted by the ions. Elemental abundances are then obtained by correcting for unseen ionic stages, as explained e.g. in  Osterbrock (1989) or Stasi\'nska (2004).

There are, however, a few issues  which have appeared over the years for which no broadly accepted solution yet exists and which cast some doubts on the derived abundances. One of them is the question of the possible presence of important ``temperature fluctuations'' in \hii\ �regions (which should better be called ``temperature inhomogeneities''). It has been argued by Peimbert (1967) that such fluctuations are likely to exist, since the temperature derived from the observed Balmer jump is significantly lower than that derived from \rOiii. Peimbert \& Costero (1969) showed that such temperature fluctuations would bias the derived abundances with respect to hydrogen towards values that are systematically too low, unless one adopts their proposed temperature fluctuation scheme to account for them. However, up to now, no mechanism has been found to produce temperature fluctuations to a level of $t^2 \sim 0.04$ which is a typical value proposed by Peimbert \& coworkers  (Peimbert 1967, Peimbert \& Peimbert 2011). It is, however, possible that several mechanisms proposed by various authors (e.g. such as those summarized by Torres-Peimbert \& Peimbert 2003) may add up to produce the observed level of temperature fluctuations.
Another, perhaps related question, is that ionic abundances derived from recombination lines are systematically higher than those derived from optical collisionally excited lines (see e.g. Garc{\'{\i}}a-Rojas \& Esteban 2007 for a review and a general discussion). These discrepancies could be due to the same temperature inhomogeneities that explain the difference between the temperatures derived from the Balmer jump and the temperatures derived from  \rOiii. In that case, it is believed that abundances obtained from recombination lines are the correct ones, since recombination lines have roughly the same -- weak -- temperature dependence, while optical collisionally excited lines have a very strong temperature dependence and are thus affected by temperature inhomogeneities (see e.g. Stasi\'nska 2009). Other interpretations of these abundance discrepancies are also possible, such as abundance inhomogeneities (Tsamis \& P\'equignot 2005, Stasi\'nska et al. 2007), density inhomogeneities (Mesa-Delgado et al. 2012), departure from a Maxwellian energy distribution of the electrons (Nicholls et al. 2012), or simply inaccurate theoretical rates for the recombination coefficients (Pradhan et al. 2011). Finally, as shown by Stasi\'nska (2005), even in the absence of abundance or density inhomogeneities, a further issue is represented by the fact that metal-rich \hii\ regions do not appear as such when using classical temperature-based abundance determinations from optical lines, due to an important internal temperature gradient caused by strong emission of infrared lines in the central zones.

In this paper, we perform a detailed analysis of the extragalactic \hii\ region H\,1013 in the spiral galaxy M 101, in order to look for some clues to the problem enunciated above. H\,1013 is a giant \hii\ region, for which deep spectra have recently been obtained on very large telescopes (Bresolin 2007 and Esteban et al. 2009). The oxygen abundances derived by these authors from optical collisionally excited lines, without accounting for temperature fluctuations, are $12\,+\,$log(O/H) $= 8.52 $ and 8.45, respectively. The abundance of \Opp\ derived by Esteban et al. (2009) from recombination lines is larger than that derived from \Oiii\ by a factor of 2.5. From this discrepancy, and attributing it to temperature fluctuations, these authors find $t^2 = 0.037 \pm 0.020$. They also obtain an estimate of $t^2$ using a maximum likelihood method from various helium line intensities and obtained $t^2 = 0.031 \pm 0.017$. On the other hand, from  his estimate of the Balmer jump temperature, $T_{\rm BJ} = 5000 \pm 800$~K, and of the temperature deduced from the \rOiii\ line ratio, $T$\oiii\ $= 7700 \pm 250$~K, Bresolin~(2007) obtains a temperature fluctuation  factor $t^2 = 0.06 \pm 0.02$, which implies an oxygen abundance larger by about a factor of 4  using the Peimbert \& Costero (1969) scheme. H\,1013 is thus an object where previous studies argued for the presence of important temperature inhomogeneities. In order to acquire additional constraints, we obtained mid-infrared spectroscopic observations with the \textit{Spitzer Space Telescope}. Mid-infrared lines are crucial in the diagnostic of the physical conditions of this object, since their emissivities depend very little on the electron temperature. In addition, they allow one to resolve the above-mentioned degeneracy in abundance determinations. In order to make the best use of this new information, we construct a photoionization model of H\,1013 that is as realistic as possible and constrained by all the available data, including the size and positions of the observing apertures. 

The organization of the paper is as follows. In Section \ref{sec:obs} we present the observational data used in this study. In Section \ref{sec:steneb} we deduce from them some properties of the ionizing radiation field and of the gas density distributions that will be used as an input in the photoionization modeling. In Section \ref{sec:photoprocedure} we describe our photoionization modeling procedure. In Section \ref{sec:results}, we present our results. A brief summary and general conclusions are presented in Section \ref{sec:conclusion}.

\section{Observational dataset}
\label{sec:obs}

\subsection{Narrow-band filter images}
\label{sec:images}

\begin{figure}[!t]
\centering
\includegraphics[angle=0,scale=0.8]{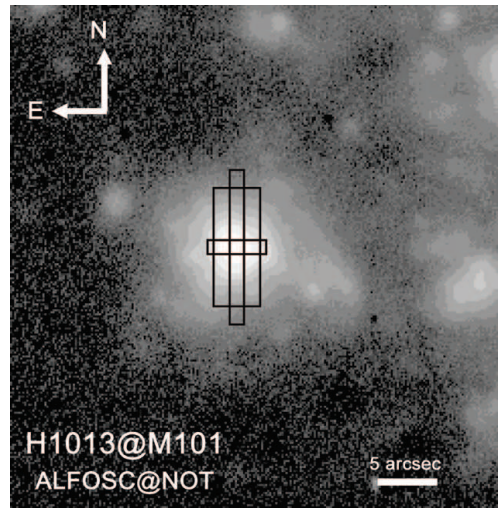}
\caption{Portion of the \Ha\ continuum-subtracted image of M\,101 
presented by Cedr\'es \& Cepa (2002) showing the \hii\ region H\,1013. The apertures 
corresponding to the optical and infrared spectra used in this study are also indicated for
reference: the long narrow one corresponds to the LRIS data, the short narrow one to the HIRES data, and the wide one to the mid-infrared  observations (note that the slit orientation is arbitrary).
\label{fig:imageslits}}
\end{figure}

 Cedr\'es \& Cepa (2002) used CDD observations in several narrow-band filters
to compile a catalogue of 338 \hii\ regions in the inner parts of M\,101 (NGC\,5457), also
providing information about their fluxes, extinctions, equivalent widths, spatial distribution, 
excitations, radiation hardness, ionization parameters and metallicities. H\,1013 is
identified as the \hii\ region number 299 in their catalogue.

We use the \Ha\ and \Hb\ continuum-subtracted images (kindly provided by B. Cedr\'es)
in our study. These images were obtained at the Nordic Optical Telescope with the ALFOSC instrument in direct imaging mode (spatial resolution
of 0.189 arcsec/pix). A description of the reduction process (also including flux 
calibration and adjacent continuum subtraction) can be found in the original paper by Cedr\'es \& Cepa.
Images through additional  narrow-band filters (\Hb, \oii, \oiii, \sii\ and \siii) are also available, however the error 
bars on the integrated line ratios  are too large ($>$50\%) to provide useful 
constraints to our models.

Figure \ref{fig:imageslits} presents a small portion of the original continuum-subtracted \Ha\ image, 
centered on H\,1013. The size and location of the apertures used to obtain the optical
and infrared spectra described in Sects. \ref{sect:optspectr} and \ref{sec:irspectr} are also indicated
for reference. As can be noticed, the three apertures cover different portions of the \hii\ region,
and none covers the whole region. This is an important point that must be taken into account when
comparing the results of our photoionization model with the various spectroscopic observations
(or when combining optical with mid-IR lines).

The continuum-subtracted \Hb\ image is used in Sect. \ref{sec:dens_struct} to obtain the nebular 
density profile. In Sect. \ref{sec:stepop} we also use the total extinction-corrected \Hb\ flux provided by Cedr\'es \& Cepa
to constrain the properties of the ionizing stellar population of H\,1013.


%

\subsection{Optical spectra}
\label{sect:optspectr}

\begin{figure}[!t]
\centering
\includegraphics[angle=90,scale=0.7]{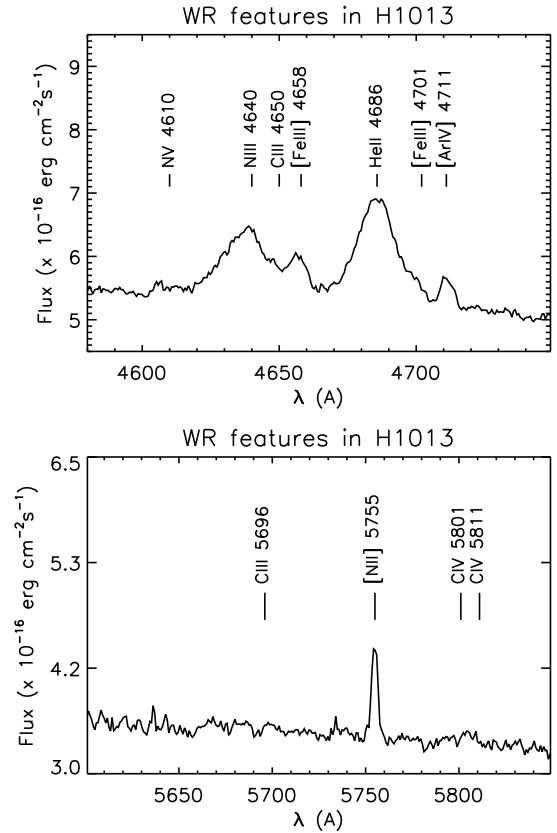}
\caption{The WR features in the optical spectrum of Bresolin (2007).
\label{fig:spectrbresolin}}
\end{figure}

Two sets of optical spectra are available. 
The first one is from 
Bresolin (2007), and was obtained with the Low Resolution Imaging Spectrometer (LRIS) spectrograph at the
Keck I telescope. The slit size was  1.5 arcsec $\times$  14 arcsec. The seeing during the observations was  0.7 arcsec. Three spectra were taken: 
a blue spectrum: 3300--5600~$\AA$ (FWHM spectral resolution $\sim$5~$\AA$); red spectrum: 4960--6670~$\AA$ (spectral resolution $\sim$ 4~$\AA$); NIR spectrum:
6050--9800~$\AA$ (spectral resolution $\sim$ 9~$\AA$). We adopt the dereddened line intensities from Bresolin (2007). 

The second spectrum we consider is that by Esteban et al (2009). It was taken with the High Resolution
Echelle Spectrometer (HIRES) at the Keck
I telescope. The spectrum covers the
3550--7440~$\AA$\ region with a spectral resolution of 
23,000. The slit width was   1.5 arcsec  and the extraction was performed over the central  5.76 arcsec. The dereddening procedure was slightly different from that adopted by Bresolin (2007) but, again, we adopted the dereddened line intensities from the original paper. 
The extraction apertures are indicated in Fig. \ref{fig:imageslits} for both spectra (note that their orientation is arbitrary). 

As a quick reference, the reddening-corrected line intensities with respect to \Hb\ are reported in the third column of Table \ref{tab:spectrmodobs}, the label in column 2 indicating whether the data are from Bresolin (2007: B07) or Esteban et al. (2009: E09). The intensities of interest from the two sets of observations agree in general very well within the quoted error bars. A notable exception is the case of \nioni{O}{ii}  3727+\footnote{Here, and in the remainder of the text, 3727+ stands for 3726 + 3729.}, whose intensity from Esteban et al. (2009) is 40\% smaller relative to the one obtained by Bresolin (2007). Such a large difference is not expected. Since the nearby high-order Balmer lines in Esteban et al. (2009) deviate from the recombination values by over two sigmas, we prefer to use the total \nioni{O}{ii}  3727+ intensity from Bresolin (2007) to constrain the photoionization models. However, we consider it safe to use the  \rOii\ line \textit{ratio} from  Esteban et al. (2009) to obtain an additional density diagnostic. Some other lines (e.g. \sii\ and \nii) have  intensities that differ by more than the assigned error bars. This, a priori, can be due to the fact that the observing apertures are not the same.

\subsection{Infrared spectrum}
\label{sec:irspectr}

\begin{figure}[!t]
\centering
\includegraphics[angle=0,scale=0.45]{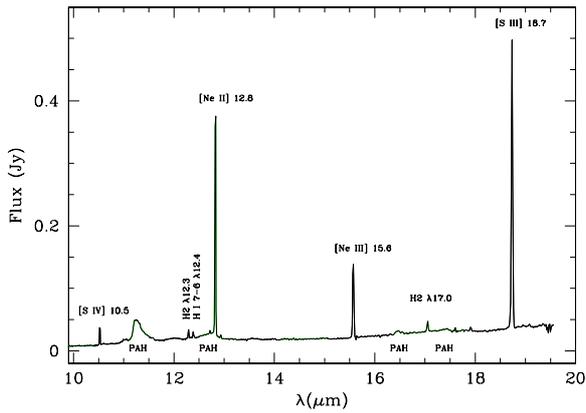}
\caption{\textit{Spitzer Space Telescope} spectrum of H\,1013.
\label{fig:spitzerspectr}}
\end{figure}

%
\begin{table}[t!]
{\scriptsize
\caption{Spitzer spectrum line measurements (Slit: 11.3 x 4.7 arcsec$^2$).
}\label{tab:spitzer}
\centering
\begin{tabular}{c c c c}
\hline \hline
\noalign{\smallskip}
Line & $\lambda$ ($\mu$m) & Flux ($\times$10$^{-14}$ erg cm$^{-2}$ s$^{-1}$) & I(\ioni{H}{i} 7-6 = 1)\\
\noalign{\smallskip}
\hline \hline
\noalign{\smallskip}
\nioni{S}{iv}    & 10.5  & 1.63$\pm$0.25  &  4.4$\pm$1.6\\ 
\ioni{H}{i} 7-6  & 12.4  & 0.37$\pm$0.12  &  1.0$\pm$0.3\\
\nioni{Ne}{ii}   & 12.8  & 15.2$\pm$1.4   &  41$\pm$14\\
\nioni{Ne}{iii}  & 15.6  & 4.95$\pm$0.20  &  13$\pm$4\\ 
\nioni{S}{iii}   & 18.7  & 12.2$\pm$0.4   &  33$\pm$11\\
\noalign{\smallskip}
\hline \hline
\end{tabular}
}
\end{table}
A mid-infrared spectrum of H\,1013 was obtained with the Infrared Spectrograph 
(IRS, Houck et al.~2004)  aboard the \textit{Spitzer Space Telescope} on 26 April 2007 (program ID 30205). The Short-High module was used to cover the 9.9-19.6 $\mu$m wavelength range with a 11.3\,$\times$\,4.7 arcsec$^2$ aperture and a spectral resolution R\,$\simeq$\,600. We acquired 13 122\,s cycles in IRS staring mode at two nod positions along the slit. The total integration time was 53 min.
The pipeline data were  processed with IRSCLEAN (version 1.9) to create bad pixels masks, and with the SPICE GUI software (version 2.0.2) for spectral extraction. 
Since  the  spectra obtained from the two nod
positions agree  within the statistical noise, we combined them  to yield the final, flux-calibrated spectrum of H\,1013, shown in 
Fig.~3.  The fluxes of the emission lines present in the spectrum were measured with the splot task within 
{\sc iraf}\footnote{{\sc iraf} is distributed by the National Optical Astronomy
Observatories, which are operated by the Association of Universities for Research in Astronomy, Inc., under cooperative agreement with the National Science Foundation.}, and are summarized in Table~1, both in terms of absolute flux (in erg\,cm$^{-2}$\,s$^{-1}$) and relative to the H~I~7-6 line.
In principle, the absolute calibration of the mid-IR data is correct  within 30\%, but we prefer to use the mid-IR and optical hydrogen lines to rescale the mid-IR spectrum (see later). This ensures that   line flux ratios are correct across the whole wavelength range.

\section{Derivation of stellar and nebular properties}
\label{sec:steneb}

\subsection{The ionizing stellar population}
\label{sec:stepop}

\begin{figure}[!t]
\centering
\includegraphics[angle=0,scale=0.45]{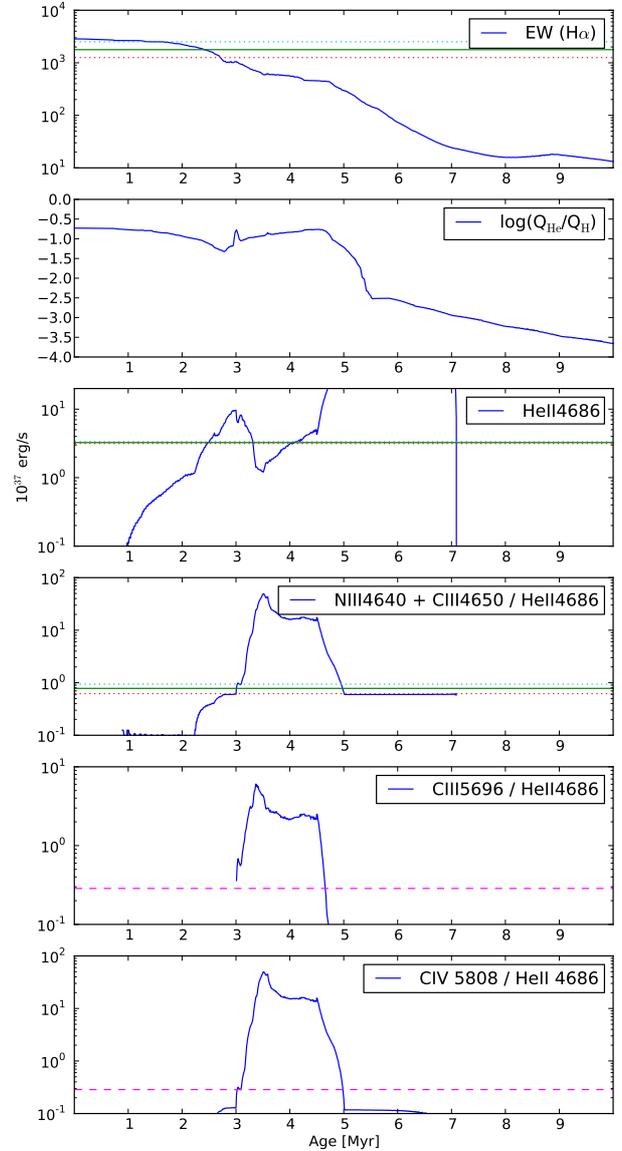}
\caption{The time dependence of several characteristics of an instantaneous starburst of solar metallicity obtained with Starburst99 (see text for details).  The observed values or their upper limits in H\,1013 are shown respectively by green and magenta horizontal lines.  \label{fig:agedet}}
\end{figure}

The first thing to note is that the spectrum of H\,1013 contains Wolf-Rayet (WR) features (see Fig. \ref{fig:spectrbresolin}). This means that, if the ionizing cluster can be approximated by an instantaneous starburst of large enough mass, its age is around 2-5 Myr. 

To obtain the total number of ionizing photons, we assume in the first place that all the stellar ionizing photons are absorbed by the nebular gas. We adopt a distance to M\,101 of 7 Mpc (a compromise between the value of  7.5 Mpc from Sandage \& Tammann 1976 and that of 6.85 Mpc from Freedman et al. 2001). From a total extinction-corrected flux in \Hb\ of $3.6 \pm 1.2 \times 10^{-13}$ erg cm$^{-2}$ s$^{-1}$ (Cedr{\'e}s \& Cepa 2002), we obtain a 
total number of hydrogen ionizing photons $Q_{\rm H} = 4.42 \times 10^{51}$s$^{-1}$. Adopting a Kroupa (2001) stellar initial mass function and using the Starburst99 stellar synthesis code (Leitherer et al. 1999) to relate stellar masses, ages and ionizing photon output, we find that, for an age of 2.5 Myr (appropriate for our object, as will be shown below),  this corresponds to a total of $\sim 1.3 \times 10^5$~\msun\ of stars with initial masses between 0.1 and 100~\msun. In fact, dust must be present inside the \hii\ region, since iron is heavily depleted (Esteban et al. 2009), and therefore absorbs part of the ionizing photons. In our final models (see below), dust absorbs about $1.5 \times 10^{51}$ ph s$^{-1}$, therefore the total initial stellar mass of the ionizing cluster is about $\sim 1.7 \times 10^5$~\msun. According to Cervi{\~n}o et al. (2003), the minimum initial cluster mass necessary to derive trustworthy results from photoionization models  at an age of about 2.5\,Myr is  at least $10^5$~\msun\ for the ions we are interested in, if using  classical synthesis models that assume full sampling of the initial mass function. Those authors however recommend a value 30 times larger  for really safe results. Clearly, our object does not comply with such conditions. In the absence of direct information on the ionizing stars (such as is sometimes available, see Jamet et al. 2004), a Monte-Carlo approach would in principle be needed. However, since the hardness of the ionizing radiation field is directly linked with the observed WR features, we may use  the stellar population synthesis code Starburst99 to estimate the spectral energy distribution that produces the  WR features observed in H 1013. 

Figure \ref{fig:agedet} shows the time dependence of several features obtained by running Starburst99 with the expanding non-LTE stellar atmosphere models implemented by Smith et al. (2002), using the Geneva tracks and high mass-loss rates. The run of Starburst99 was made for solar metallicity, which is the most appropriate for our object. From top to bottom, Fig. \ref{fig:agedet} shows the values of 
\begin{enumerate}
 \item  EW(\Ha), the equivalent width of the \Ha\ emission for an instantaneous burst,  
  \item the ratio  $Q_{\rm He}$/$Q_{\rm H}$ of helium to hydrogen ionizing photons predicted by the model,
  \item the intensity of the stellar  He\,{\sc ii}\,$\lambda$4686 feature ($2.5 \times 10^{-15}$ erg cm$^{-2}$ s$^{-1}$),
    \item the intensity of the stellar N\,{\sc iii}\,$\lambda$4640 + C\,{\sc iii}\,$\lambda$4650 feature with respect to He\,{\sc ii}\,$\lambda$4686,
  \item the intensity of the stellar C\,{\sc iii}\,$\lambda$5696  feature with respect to He\,{\sc ii}\,$\lambda$4686,
    \item and the intensity of the stellar C\,{\sc iv}\,$\lambda$5808  feature with respect to He\,{\sc ii}\,$\lambda$4686 . 

\end{enumerate}

The observed values are represented with the green horizontal lines, the error bars with red dotted lines, and the upper limits (for C\,{\sc iii}\,$\lambda$5696 and C\,{\sc iv}\,$\lambda$5808) are represented with magenta lines.  The $Q_{\rm He}$/$Q_{\rm H}$ ratio is actually not observed, but this value determines the observed value of the emission line ratio \Oiii/\Oii\ in the nebula. 
From Fig. \ref{fig:agedet}, we see that the observed equivalent width of the \Ha\ line indicates an age of 2--3 Myr, the intensity of the stellar  He\,{\sc ii}\,$\lambda$4686 feature is compatible with ages of about 2.5, 3.2 and 4.2 Myr, while the  N\,{\sc iii}\,$\lambda$4640 + C\,{\sc iii}\,$\lambda$4650 feature indicates an age of 2.5--3 Myr, or larger than 5 Myr. On the other hand, the absence of observed stellar carbon features indicates an age smaller than 3 Myr. In fact, in the framework of an instantaneous starburst, the observational constraints on the stellar population are almost incompatible, but taking into account uncertainties, they point to an age of  2.5 Myr. However, if we consider that, in practice,  a starburst is never instantaneous but occurs during a time interval of a few $10^5$ yr, the range of possible solutions is larger.  We approximate such a situation by considering two instantaneous starbursts of different intensities separated by a small time interval.   After trials and errors, we find that combining one burst giving 
$Q_{\rm H}=4.6 \times 10^{51}$ ph s$^{-1}$ 
at an age of $2.5 \times 10^6$ yr and another one giving 
$Q_{\rm H}=1 \times 10^{50}$ ph s$^{-1}$ 
at an age of $3 \times 10^6$ yr correctly reproduces the observational constraints on the stellar population\footnote{This combination of ages does not have to correspond to reality since, as explained above, we are in the regime where statistical fluctuations matter. The important thing is that the spectral energy distribution of the ionizing radiation field is correct.} and allows us to correctly reproduce the observed emission line ratios of the nebula (see below).

\subsection{Simple plasma diagnostics}
\label{fsec:diagn}

\begin{figure}[!t]
\centering
\includegraphics[angle=0,scale=0.55]{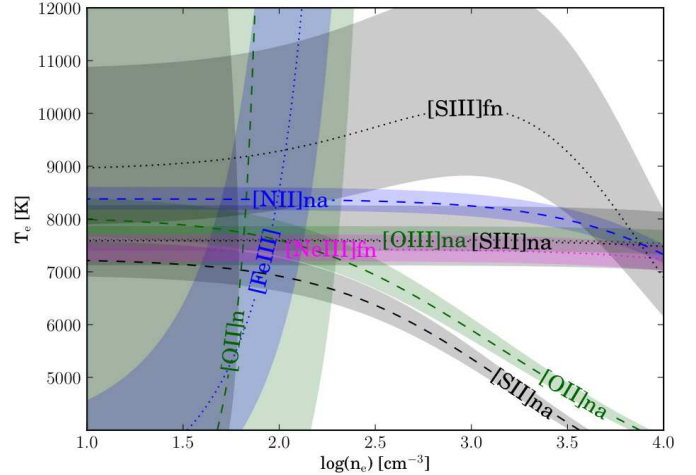}
\caption{Plasma diagnostics using PyNeb. The dashed or dotted values indicate the density-temperature loci corresponding to the observed reddening-corrected line ratios (also aperture-corrected when combining \textit{Spitzer} and optical data, as explained in Sect. \ref{sec:photoprocedure}). The coloured bands delineate the one sigma error zones in the corresponding ratios. Diagnostics involving oxygen ions are in green:  [O\,{\sc{ii}}]na  3727+/7325+; [O\,{\sc{ii}}]n: 3729/3726; [O\,{\sc{iii}}]na: 5007/4363. Diagnostics involving sulfur ions are in black: [S\,{\sc{ii}}]n:  6716/6731; [S\,{\sc{ii}}]na: 6720+/4072+; [S\,{\sc{iii}}]fn: 18.7$\mu$m/9069; [S\,{\sc{iii}}]na: 9069/6312; the [N\,{\sc{ii}}]na 6584/5755 diagnostic  is in blue; the [Ne\,{\sc{iii}}]fn 15.6$\mu$m/3869 diagnostic  is in pink; finally the  [Fe\,{\sc{iii}}] 5272/4987 diagnostic is in blue.  (The characters n, a, f in the denomination of the line ratios stand for `nebular', `auroral' and `fine-structure', respectively.)    \label{fig:pyneb}}
\end{figure}




Before proceeding to the description of the photoionization modeling, it is useful to consider the temperature-density diagnostic diagram for H\,1013. This diagram, presented in Fig.  \ref{fig:pyneb}, 
  is obtained by considering all the observed temperature and/or density sensitive intensity ratios of lines of the same ion that have been observed in H\,1013. We use the extinction-corrected intensity measurements and associated uncertainties from Bresolin~(2007), except in the case of the \rOii\ ratio, which was not available in those observations and is taken, together with the [Fe\,{\sc{iii}}] $\lambda$ 5272/4987 line ratio, from Esteban et al.~(2009). Note that line ratios involving one optical and one infrared line are obtained by forcing the  H 7-6 / \Hb\ ratio to match the case B recombination value  of 0.0098 for a temperature of 10000~K and a density of 100 cm$^{-3}$ (Storey \& Hummer 1995), and the uncertainties are dominated by the uncertainty in the flux of the H 7-6 line. The diagram has been constructed with the software PyNeb  (Luridiana et al. 2012) using the same atomic data  as in  the photoionization Cloudy (Ferland et al. 1998) in the version adopted for the present study. This diagram is of course only indicative, since, for each line ratio, it assumes that both density and temperature are constant.
Note that the effects of recombinations are not accounted for in such a diagram, while they are taken into account in Cloudy. In the present case, this could affect the [O\,{\sc{ii}}]  $\lambda$3727+/7325+ ratio. Another approximation is that, when matching the optical and infrared spectra using the hydrogen lines, one ignores the ionization structure that affects the \neiii\ and \siii\ lines, which are observed both in the optical and in the infrared ranges, but through different apertures.

In addition to the diagnostics obtained by Bresolin (2007) and Esteban et al. (2009), [Fe\,{\sc{iii}}] $\lambda$5272/4987 provides a useful diagnostic, as it restricts the density in the corresponding emitting zone to a value roughly below 160 cm$^{-3}$. The [S\,{\sc{ii}}] $\lambda$6716/6731 diagnostic indicates an upper limit of the density of 50 cm$^{-3}$ in the \sii\ emitting zone. With the \textit{Spitzer Space Telescope} observations, we have two additional temperature diagnostics, that could provide some clues on the possible presence of large temperature inhomogeneities. If we adopt the Peimbert (1967) temperature fluctuation scheme, the values of  $T_0=5500$~K and $t^2 = 0.06$ inferred by Bresolin (2007) would imply a temperature derived from the  [Ne\,{\sc{iii}}] $\lambda$15.6$\mu$m/3869 ratio of 6150~K, and a temperature derived from the [S\,{\sc{iii}}] $\lambda$18.7$\mu$m/9069  ratio of 5500~K. A value of $t^2 = 0.03 - 0.04$, as derived by Esteban et al. (2009)  would imply  a [Ne\,{\sc{iii}}] $\lambda$15.6$\mu$m/3869 temperature  of $6400-6300$~K, and a  [S\,{\sc{iii}}] $\lambda$18.7$\mu$m/9069  temperature of $6100-5900$~K. Such values are well below the ones obtained from the \textit{Spitzer} measurements, which are roughly $7000\pm200$~K and $9200^{+ 2000}_{-1000}$~K, as inferred respectively from  Fig. \ref{fig:pyneb}.  

While the diagram presented in  Fig.  \ref{fig:pyneb} is useful to visualize the various temperature and density diagnostics, it includes several approximations, as mentioned above. The only way to account for all the observational constraints in the most accurate way is to build a tailored photoionization model that is able to reproduce \textit{each} of the observational constraints within the uncertainties. 


\subsection{Global density structure}
\label{sec:dens_struct}

\begin{figure}[!t]
\centering
\includegraphics[angle=0,scale=0.4]{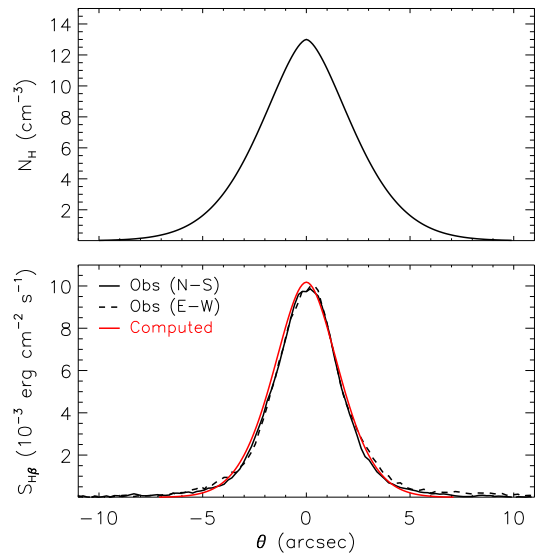}
\caption{Upper panel: Density law needed to fit the   \Hb\ surface brightness distribution assuming 
a spherical geometry. Lower panel: The observed \Hb\ surface brightness distribution 
in two perpendicular directions (black continuous and dotted lines) and the computed one assuming a constant electron temperature. 
\label{fig:density_law}}
\end{figure}
The first requirement for a successful photoionization model is to be able to reproduce the observed \Hb\ surface brightness distribution. 
 We use the \Hb\ (extinction corrected, continuum subtracted, and flux calibrated) image from Cedr\'es \& Cepa (2002) and assume a uniform electron temperature in order to obtain the surface brightness distribution in two perpendicular directions. The two distributions are quite similar, indicating that the nebula can be assumed to be spherically symmetric. We use this \Hb\ surface brightness distribution to obtain the density law. We
assume  spherical symmetry and find that a density law of the type: 
\begin{equation}
\label{ed:distrdens}
N_{\rm H}=A_1 \exp (-(r/A_2)^{A_3}) 
\end{equation} 
 is convenient to reproduce the observed \Hb\ surface brightness distribution assuming a constant electron temperature.
 We find a good fit  with A$_{1}$=13 cm$^{-3}$, A$_{2}$=107 pc, A$_{3}$=1.6, as shown in Fig. \ref{fig:density_law}.
Note that in this case we do not consider any filling factor. In the models, the value of $A_1$ is adjusted to reproduce the density and ionization diagnostics, and a filling factor smaller than unity is then needed to reproduce the observed nebular size.  
In the process of our model fitting, we check that the computed radial temperature variation has not drastically modified the  \Hb\ surface brightness distribution.

\section{Photoionization modelling procedure}
\label{sec:photoprocedure}

%
\begin{table}[t!]
{\scriptsize
\caption{Line ratios used to constrain the model and the associated tolerances.}
\label{tab:Tolerance}
\centering
\begin{tabular}{c c}
\hline \hline
\noalign{\smallskip}
$\Delta  O/O$ & line ratio \\
\hline \hline
\noalign{\smallskip}
 \noalign{\smallskip}
\multicolumn{2}{c}{density indicators}\\
 \noalign{\smallskip} 
 0.15	&       \nioni{O}{II}3726 / \nioni{O}{II}3729  \\
 0.40       & \nioni{Fe}{III}4702 / \nioni{Fe}{III}4659 \\
 0.40       & \nioni{Fe}{III}4659 / \nioni{Fe}{III}5271 \\
  0.15	&       \nioni{S}{II}6731 / \nioni{S}{II}6716  \\ 
  0.15	&   \nioni{Cl}{III}5538 / \nioni{Cl}{III}5518  \\
 \noalign{\smallskip}
\multicolumn{2}{c}{temperature indicators}\\
 \noalign{\smallskip}
 0.30	&   \nioni{Ne}{III}3869 / \nioni{Ne}{III}15.5  \\
 0.12	&     \nioni{O}{III}4363 / \nioni{O}{III}5007  \\
 0.30	&     \nioni{S}{III}6312 / \nioni{S}{III}9069  \\
0.40	        &     \nioni{S}{III}9069 / \nioni{S}{III}18.7  \\
 0.10	&       \nioni{N}{II}5755 / \nioni{N}{II}6584  \\
 0.10	&     \nioni{O}{II}7325+ / \nioni{O}{II}3727+  \\
 0.10	&     \nioni{S}{II}4070+ / \nioni{S}{II}6716+  \\
 \noalign{\smallskip}
\multicolumn{2}{c}{ionization structure}\\
 \noalign{\smallskip}
 0.30   &      \nioni{S}{III}9069 / \nioni{S}{II}6716+ \\  
 0.40	&      \nioni{S}{IV}10.5 / \nioni{S}{III}18.7 \\ 
 0.15	&     \nioni{O}{III}5007 / \nioni{O}{II}3727+ \\ 
 0.40	&    \nioni{Ne}{III}15.5 / \nioni{Ne}{II}12.8 \\ 
   \noalign{\smallskip}
\multicolumn{2}{c}{abundances}\\
 \noalign{\smallskip}
 0.05   &  	     HeI 5876 / H$\alpha$  \\
 0.06	&  	   \nioni{N}{II}6584 / H$\alpha$ \\ 
 0.05	&  	  \nioni{O}{III}5007 / H$\beta$ \\ 
 0.06	&  	   \nioni{S}{II}6716+ / H$\alpha$ \\ 
 0.08	&  	 \nioni{Ne}{III}3869 / H$\beta$ \\ 
 0.08	&  	 \nioni{Cl}{III}5525+ / H$\beta$ \\ 
0.08	&  	 \nioni{Ar}{III}7135 / H$\alpha$ \\ 
0.15	&  	 \nioni{Fe}{III}4659 / H$\alpha$ \\ 
   \noalign{\smallskip}
\hline \hline
\end{tabular}
}
\end{table}

The photoionization modeling is performed using the code Cloudy,  version c08.01 (Ferland et al.~1998), within the pyCloudy\footnote{pyCloudy is the python version of Cloudy\_3D (Morisset, 2006) available at https://sites.google.com/site/pycloudy/} environment. With pyCloudy, we can easily obtain the line intensities in specific apertures. The nebula is assumed to be spherically symmetric, with the density distribution described in Sect. \ref{sec:dens_struct}.  The stellar ionizing radiation field is obtained as described in Sect. \ref{sec:stepop}. The computed line intensities are compared with the observed ones, each in its corresponding aperture. The effects of seeing for the optical data and  of minor irregularities in the nebular density distribution are taken into account by convolving with a gaussian  (adopting a  width of 1 arcsec for the optical lines and 0.5 arcsec for the infrared lines).

The initial chemical composition of the gas is that derived from collisionally excited lines (and taking $t^2 = 0$) by Bresolin (2007) for O, N, and Ne and by Esteban et al. (2009) for He, S, Cl, Ar, Fe, which were not given by Bresolin. The reason why we use the O, Ne, Ne abundances by Bresolin is due to the potential problem in the \Oii\ intensity of Esteban et al. (2009) as explained in Sect. \ref{sect:optspectr}. 
The carbon abundance is taken equal to the value derived by Bresolin (2007) from the \ioni{C}{ii }$\lambda$4267 recombination line, since there is no other direct information on the carbon abundance. 

In the classical logarithmic units taking 12 for hydrogen, the initial composition of the gas is thus: H$=12$,  He$=10.87$,  C$=8.66$,  N$=7.57$, O$=8.52$,  Ne$=7.41$,  S$=6.91$,   Cl$=5.50$,  Ar$=6.35$, Fe$=5.74$. 

Since iron is observed to be strongly depleted in this object, we consider the presence of dust in the model, using the ``grains ism'' option of Cloudy (but see below for the radial distribution of the grains). In the present object, the effect of grains is essentially to absorb part of the ionizing photons (about 30\%) and to heat the gas, especially in the central zone.

A satisfactory model must reproduce the observed \Hb\ surface brightness distribution law (which in our case it does by construction), the total reddening-corrected \Hb\ flux and the observed nebular angular size, as well as each of the reddening-corrected observed line ratios in each slit, within the observational uncertainties. We emphasize that it is not sufficient to perform a $\chi^2$ fitting of the sum of the deviation of line intensities to the observed value: each line carries its own information, and it is important that all observational constraints be properly accounted for. 

Absolute calibration of spectroscopic observations is difficult. We intercalibrate the \textit{Spitzer} and optical data by forcing the measured value of the H 7-6 / \Hb\  ratio to the one predicted by our photoionization models in the corresponding slits. The value of $f$(IR), representing the factor by which the measured IR fluxes have to be multiplied in order for the H 7-6 / \Hb\  ratio to be in agreement with the model, turns out to be  $\sim$1.07.
To judge a model, we adopt the same policy as in Morisset \& Georgiev (2009) and Stasi\'nska et al. (2010). For each observable $O$ we fix a tolerance $\tau(O)$, given by  
\begin {equation} 
   \tau(O) = {\rm log} (1+\Delta  O/O),
\end {equation}
where $\Delta  O/O$ is the maximum ``acceptable'' deviation. 
The values  adopted for $\tau(O)$  take into account the observational uncertainties and are chosen according to our experience with model fitting. For example, while the observational uncertainty in  \Oiii/\Oii\ is small, we take a tolerance of 1.15 to reflect the fact that this line ratio is very sensitive to the density distribution and/or the slit position in our object. 

It is convenient to divide the line ratios to be fitted in different categories:
\begin{itemize}
  \item Ratios of lines measuring the density.
    \item Ratios of lines measuring the electron temperature.
  \item Ratios of hydrogen lines or of helium lines, which test the reddening correction, among others.
    \item Ratios of two different ions of the same element, testing  the ionization structure. 
  \item Ratios of lines used to determine the chemical composition, once the ionization structure is correct.
  \end{itemize}
  
 The model is compared to the observations by looking at the values of 
\begin {equation} 
  \kappa(O) = ({\rm log} O_{\rm mod} - {\rm log} O_{\rm obs})/\tau(O),
\end {equation} 
which can easily be plotted in the same diagram for all the line ratios. The list of line ratios  and associated values of $\Delta  O/O$ is
given in  Table \ref{tab:Tolerance}.  

During the fitting procedure, we vary the values of A$_{1}$, the parameter defining the density (see Sect. \ref{sec:dens_struct}), the filling factor, the total number of ionizing photons (since the proportion of those photons that are absorbed by dust is not known a priori), and the elemental abundances. We also had to modify the spatial distribution of the internal dust. Indeed, when considering a uniform dust-to-gas ratio in the entire nebula, the inner zones became too hot, and the \rOiii\ ratio could not be fitted to gether with the temperature diagnostics of the low excitation zones. We adopted the simplest model for the dust-to-gas ratio, which is a step function. The parameters of the step function were obtained by trial and error to reproduce the observed temperature constraints. Note that such a dust distribution which may seem \textit{ad hoc} is actually fully justified by the effects of  radiation pressure  and the possible melting of grains close to the star. That this is precisely what occurs near the Trapezium stars in the Orion nebula was shown already long ago (Baade \& Minkowski 1937).

\section{Results}
\label{sec:results}

\begin{figure*}[!t]
\centering
\includegraphics[angle=0,scale=0.4]{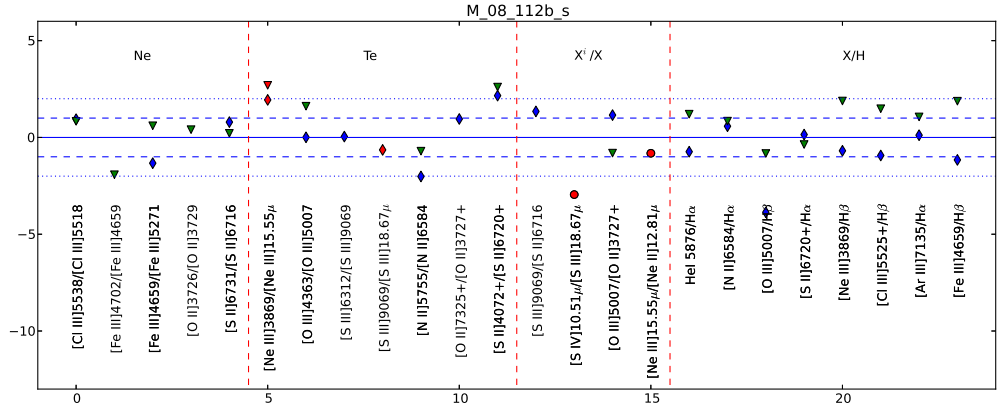}
\caption{Comparison of model M 1 with the observations. Blue diamonds represent data from Bresolin (2007), green triangles data from Esteban et al. (2009). Red symbols represent ratios involving an infrared line.   The dashed and dotted lines represent 1 sigma and 2 sigmas deviations, respectively. \label{fig:compar1}}
\end{figure*}

\begin{figure*}[!t]
\centering
\includegraphics[angle=0,scale=0.4]{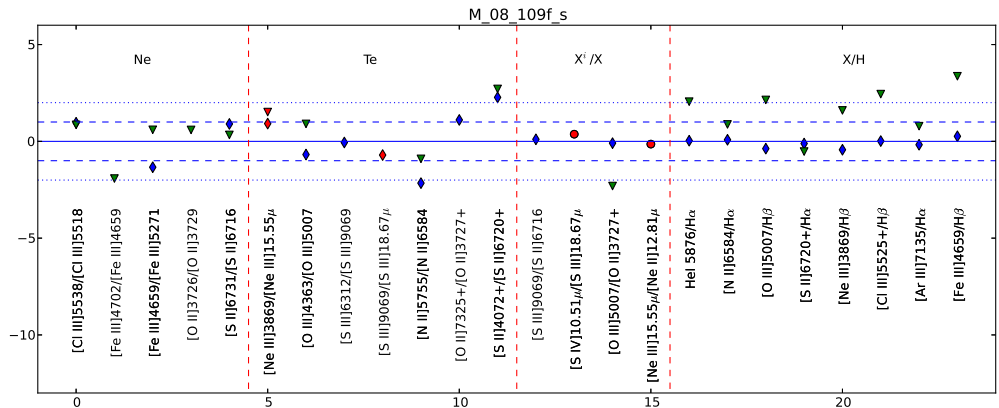}
\caption{Comparison of the composite model, M 2,  with the observations. Same layout as in Fig. \ref{fig:compar1}.
\label{fig:compar2}}
\end{figure*}


\begin{table*}[t!]
{\scriptsize
\caption{Observed line ratios in H 1013 compared to the results of models M1 and M2.  }
\label{tab:spectrmodobs}
\centering
\begin{tabular}{llrrr}
\hline \hline
\noalign{\smallskip}
 	line intensities	$^a$										 &	references &  observed  &		 mod. M1  & mod. M2    \\
\hline \hline
\noalign{\smallskip}
\nioni{O}{ii} 3726                 	 	 	 	 & 	 	E09 	 & 	   57.00  $\pm$     2.00 	 & 	63.02	 & 	91.17	  \\
\nioni{O}{ii} 3729                 	 	 	 	 & 	 	E09 	 & 	   78.00  $\pm$     3.00 	 & 	81.51	 & 	114.86	  \\
\nioni{O}{ii} 3727+                 			 	 & 	  {\color{blue}	B07}  	 & 	  221.00  $\pm$    11.00 	 & 	155.71	 & 	219.82	  \\
\nioni{O}{ii} 3727+                			 	 & 	 	E09 	 & 	  135.00  $\pm$     5.00 	 & 	145.16	 & 	206.9	  \\
H13 3734                   	 				 	 & 	 	E09 	 & 	    1.50  $\pm$     0.10 	 & 	2.29	 & 	2.28	  \\
H12 3750                   	 				 	 & 	 	 {\color{blue}B07}  	 & 	    3.00  $\pm$     0.20 	 & 	2.92	 & 	2.9	  \\
H12 3750                   	 				 	 & 	 	E09 	 & 	    2.00  $\pm$     0.20 	 & 	2.91	 & 	2.89	  \\
H11 3771                   					 	 &  	 {\color{blue} 	B07 } 	 &  	    3.30  $\pm$     0.20 	 &  	3.81	 &  	3.79	  \\
H11 3771                   					 	 & 	 	E09 	 & 	    3.10  $\pm$     0.20 	 & 	3.80	 & 	3.77	  \\
H10 3798                   					 	 & 	  {\color{blue}	B07}  	 & 	    4.40  $\pm$     0.20 	 & 	5.11	 & 	5.08	  \\
H10 3798                   	 				 	 & 	 	E09 	 & 	    4.20  $\pm$     0.20 	 & 	5.10	 & 	5.07	  \\
H9 3835                    	 				 	 & 	  {\color{blue}	B07}  	 & 	    7.30  $\pm$     0.30 	 & 	7.09	 & 	7.05	  \\
H9 3835                    	 				 	 & 	 	E09 	 & 	    6.40  $\pm$     0.30 	 & 	7.07	 & 	7.03	  \\
\nioni{Ne}{iii} 3869                 	 & 	  {\color{blue}	B07}  	 & 	    3.10  $\pm$     0.20 	 & 	2.94	 & 	3	  \\
\nioni{Ne}{iii} 3869                 	 & 	 	E09 	 & 	    2.80  $\pm$     0.20 	 & 	3.24	 & 	3.17	  \\
H7 3970                    	 				 	 &  	 	 {\color{blue}B07}  	 &  	   16.40  $\pm$     0.70 	 &  	15.57	 &  	15.52	  \\
H7 3970                    	 				 	 & 	 	E09 	 & 	   15.80  $\pm$     0.60 	 & 	15.55	 & 	15.5	  \\
 \ioni{He}{i} 4026                   	 	 	 & 	  {\color{blue}	B07}  	 & 	    1.54  $\pm$     0.08 	 & 	1.91	 & 	1.98	  \\
 \ioni{He}{i} 4026                   	 	 	 & 	 	E09 	 & 	    1.70  $\pm$     0.10 	 & 	2.04	 & 	2.12	  \\
\nioni{S}{ii}  4072+                	 	 	 	 & 	  {\color{blue}	B07}  	 & 	    1.27  $\pm$     0.07 	 & 	1.58	 & 	1.58	  \\
\nioni{S}{ii}  4072+                	 	 	 	 & 	 	E09 	 & 	    1.06  $\pm$     0.18 	 & 	1.39	 & 	1.39	  \\
H$\delta$                         	 				 	 & 	 	 {\color{blue}B07}  	 & 	   26.20  $\pm$     1.10 	 & 	25.48	 & 	25.43	  \\
H$\delta$                         	 				 	 & 	 	E09 	 & 	   25.30  $\pm$     0.90 	 & 	25.47	 & 	25.41	  \\
\ioni{C}{ii }4267                  	 	 	 	 &  	  {\color{blue}	B07}  	 &  	    0.25  $\pm$     0.03 	 &  	0.30	 &  	0.28	  \\
\ioni{C}{ii} 4267                  	 	 	 	 & 	 	E09 	 & 	    0.28  $\pm$     0.08 	 & 	0.33	 & 	0.31	  \\
H$\gamma$                         	 				 	 & 	  {\color{blue}	B07}  	 & 	   46.40  $\pm$     1.90 	 & 	46.37	 & 	46.32	  \\
H$\gamma$                         	 				 	 & 	 	E09 	 & 	   44.00  $\pm$     1.00 	 & 	46.35	 & 	46.3	  \\
   \nioni{O}{iii} 4363                  	 & 	  {\color{blue}	B07}  	 & 	    0.24  $\pm$     0.03 	 & 	0.20	 & 	0.22	  \\
   \nioni{O}{iii} 4363                  	 & 	 	E09 	 & 	    0.19  $\pm$     0.07 	 & 	0.22	 & 	0.23	  \\
 \ioni{He}{i} 4388                   	 	 	 & 	  {\color{blue}	B07}  	 & 	    0.40  $\pm$     0.05 	 & 	0.51	 & 	0.53	  \\
 \ioni{He}{i} 4388                   	 	 	 & 	 	E09 	 & 	    0.52  $\pm$     0.09 	 & 	0.55	 & 	0.57	  \\
 \ioni{He}{i} 4471                   	 	 	 &  	  {\color{blue}	B07}  	 &  	    3.90  $\pm$     0.20 	 &  	4.10	 &  	4.26	  \\
\ioni{O}{ii} 4651  $^b$                	 	 	 	 & 	 	E09 	 & 	    0.62  $\pm$     0.27 	 & 	0.12	 & 	0.15	  \\
\nioni{Fe}{iii} 4659                 	 & 	  {\color{blue}	B07}  	 & 	    0.40  $\pm$     0.05 	 & 	0.34	 & 	0.41	  \\
\nioni{Fe}{iii} 4659                 	 & 	 	E09 	 & 	    0.25  $\pm$     0.04 	 & 	0.32	 & 	0.4	  \\
\nioni{Fe}{iii} 4702                 	 & 	 	E09 	 & 	    0.14  $\pm$     0.03 	 & 	0.10	 & 	0.12	  \\
\nioni{Fe}{iii} 4987                 	 & 	  {\color{blue}	B07}  	 & 	    0.35  $\pm$     0.04 	 & 	0.01	 & 	0.01	  \\
\nioni{Fe}{iii} 4987                 	 & 	 	E09 	 & 	    0.32  $\pm$     0.04 	 & 	0.01	 & 	0.01	  \\
   \nioni{O}{iii} 5007                  	 & 	 	 {\color{blue}B07}  	 & 	  102.00  $\pm$     4.00 	 & 	84.46	 & 	100.15	  \\
   \nioni{O}{iii} 5007                  	 &  	 	E09 	 &  	   97.00  $\pm$     3.00 	 &  	93.16	 &  	107.68	  \\
\nioni{Fe}{iii} 5271                 	 & 	 	 {\color{blue}B07}  	 & 	    0.15  $\pm$     0.02 	 & 	0.20	 & 	0.24	  \\
\nioni{Fe}{iii} 5271                 	 & 	 	E09 	 & 	    0.18  $\pm$     0.04 	 & 	0.19	 & 	0.24	  \\
\nioni{Cl}{iii} 5518                 	 & 	 	 {\color{blue}B07}  	 & 	    0.46  $\pm$     0.03 	 & 	0.41	 & 	0.44	  \\
\nioni{Cl}{iii} 5518                 	 & 	 	E09 	 & 	    0.39  $\pm$     0.05 	 & 	0.42	 & 	0.45	  \\
\nioni{Cl}{iii} 5538                 	 & 	 	 {\color{blue}B07}  	 & 	    0.29  $\pm$     0.02 	 & 	0.29	 & 	0.31	  \\
\nioni{Cl}{iii} 5538                 	 & 	 	E09 	 & 	    0.25  $\pm$     0.04 	 & 	0.30	 & 	0.32	  \\
\nioni{N}{ii} 5755                 	 	 	 	 & 	 	 {\color{blue}B07}  	 & 	    0.59  $\pm$     0.03 	 & 	0.51	 & 	0.49	  \\
\nioni{N}{ii} 5755                 	 	 	 	 &  	 	E09 	 &  	    0.43  $\pm$     0.05 	 &  	0.44	 &  	0.43	  \\
 \ioni{He}{i} 5876                   	 		 & 	 	 {\color{blue}B07}  	 & 	   12.10  $\pm$     0.60 	 & 	11.74	 & 	12.21	  \\
 \ioni{He}{i} 5876                   	 		 & 	 	E09 	 & 	   11.30  $\pm$     0.50 	 & 	12.52	 & 	13.07	  \\
   \nioni{O}{i} 6300                  	 	 	 	 & 	  {\color{blue}	B07}  	 & 	    1.21  $\pm$     0.07 	 & 	0.16	 & 	0.35	  \\
   \nioni{O}{i} 6300                  	 	 	 	 & 	 	E09 	 & 	    1.44  $\pm$     0.09 	 & 	0.12	 & 	0.27	  \\
\nioni{S}{iii} 6312                  	 & 	 	 {\color{blue}B07}  	 & 	    0.77  $\pm$     0.05 	 & 	1.07	 & 	0.74	  \\
\nioni{S}{iii} 6312                  	 & 	 	E09 	 & 	    0.72  $\pm$     0.06 	 & 	1.08	 & 	0.74	  \\
H$\alpha$                         	 				 	 & 	 	 {\color{blue}B07}  	 & 	  293.00  $\pm$    16.00 	 & 	294.77	 & 	295.06	  \\
H$\alpha$                         	 				 	 &  	 	E09 	 &  	  282.00  $\pm$    16.00 	 &  	294.71	 &  	295.04	  \\
\nioni{N}{ii} 6584                 	 	 	 	 & 	 	 {\color{blue}B07}  	 & 	   71.00  $\pm$     4.00 	 & 	73.88	 & 	71.84	  \\
\nioni{N}{ii} 6584                 	 	 	 	 & 	 	E09 	 & 	   59.00  $\pm$     3.00 	 & 	64.77	 & 	64.96	  \\
 \ioni{He}{i} 6678                   	 	 	 & 	  {\color{blue}	B07}  	 & 	    3.40  $\pm$     0.20 	 & 	3.34	 & 	3.47	  \\
 \ioni{He}{i} 6678                   	 	 	 & 	 	E09 	 & 	    3.10  $\pm$     0.20 	 & 	3.56	 & 	3.72	  \\
\nioni{S}{ii}  6716                 	 	 	 	 & 	  {\color{blue}	B07}  	 & 	   17.80  $\pm$     1.00 	 & 	17.25	 & 	16.9	  \\
\nioni{S}{ii} 6716                 	 	 	 	 & 	 	E09 	 & 	   14.90  $\pm$     0.90 	 & 	15.05	 & 	14.82	  \\
\nioni{S}{ii}  6731                 	 	 	 	 & 	 	 {\color{blue}B07}  	 & 	   12.00  $\pm$     0.70 	 & 	12.99	 & 	12.92	  \\
\nioni{S}{ii}  6731                 	 	 	 	 &  	 	E09 	 &  	   11.00  $\pm$     0.70 	 &  	11.45	 &  	11.46	  \\
\nioni{Ar}{iii} 7135                 	 & 	 	 {\color{blue}B07}  	 & 	    7.10  $\pm$     0.40 	 & 	7.21	 & 	7.05	  \\
\nioni{Ar}{iii} 7135                 	 & 	 	E09 	 & 	    6.80  $\pm$     0.50 	 & 	7.71	 & 	7.56	  \\
   \nioni{O}{ii} 7325+                	 	 	 	 & 	 	 {\color{blue}B07}  	 & 	    2.40  $\pm$     0.20 	 & 	1.85	 & 	2.65	  \\
\nioni{S}{iii} 9069                  	 & 	  {\color{blue}	B07}  	 & 	   24.00  $\pm$     5.00 	 & 	33.10	 & 	23.42	  \\
\nioni{S}{iV} 10.51$\mu$m    				 	 & 	  IR  		 & 	    4.40  $\pm$     1.94 	 & 	2.81	 & 	6.2	  \\
\nioni{Ne}{ii} 12.81$\mu$m   	 	 	 & 	  IR  		 & 	   41.00  $\pm$    17.01 	 & 	30.92	 & 	32.84	  \\
\nioni{Ne}{iii} 15.55$\mu$m  	 & 	  IR  		 & 	   13.00  $\pm$     4.86 	 & 	7.45	 & 	9.93	  \\
\nioni{S}{iii}  18.67$\mu$m   	 &  	  IR  		 &  	   33.00  $\pm$    13.36 	 &  	56.81	 &  	41.02	  \\
log L(H$\alpha$)   [erg s$^{-1}$]                  				 	 & 	          	 & 	39.78	 & 	39.78	 & 	39.78	  \\
\noalign{\smallskip}
\hline \hline
\multicolumn{5}{l}{$^a$ In units of H$\beta=100$ for optical lines and of  H7-6 12.4$\mu$m$=1$ for infrared lines} \\
\multicolumn{5}{l}{$^b$ Represents the sum of the multiplet 1 of \ion{O}{ii}} \\
\end{tabular}
}
\end{table*}

\subsection{The simplest model}
\label{sec:simplest}

In order to constrain the model-fitting  we took the observations by Bresolin (2007), using the data of Esteban et al. (2009) as a complement. 
The most satisfactory model we produced, within the explained procedure, is one having  $A_1 = 200$ cm$^{-3}$, a filling factor of 0.005, and a total number of ionizing photons $Q$(H)\,=\,5.9$\times10^{51}$s$^{-1}$.

A graphical representation of how the line-ratio constraints  match the data is shown in Fig. \ref{fig:compar1}: the values of $\kappa(O)$ are
plotted for all the constraints $O$. Diamonds correspond to intensities from the Bresolin (2007) spectrum, triangles to intensities from the Esteban et al. (2009) optical spectrum, each one within its extraction aperture. It can be seen that \textit{each} of the
values of $\kappa(O)$ is  found to lie between $-2$ and $+2$ (except for the \nioni{S}{iv} $\lambda$10.5$\mu$m/\nioni{S}{iii} $\lambda$18.7$\mu$m ratio, and the \Oiii/\Hb\ ratio from Bresolin 2007). A more conventional comparison of our best model, M1,  with the observed line ratios is shown in Table \ref{tab:spectrmodobs}. 
It can be seen that the $\kappa(O)$ values for Bresolin (2007) and Esteban et al. (2009) are slightly different, which means that either the density distribution in the model is too simplistic (i.e. that  some slight density inhomogeneities are present) or that the observational error bars are slightly underestimated. Probably both reasons are relevant, but in any case the differences are small (except for the case of the \Oii\ line mentioned above).

The elemental abundances in this model are:  H$=12$,  He$=10.97$,  C$=8.66$,  N$=7.79$, O$=8.42$,  Ne$=7.80$,  S$=6.98$,  Cl$=5.18$,  Ar$=6.20$, Fe$=5.77$.  
They differ from the initial ones, mainly due to the fact that the classical ionization correction factors used to derive abundances by Bresolin (2007) or Esteban et al. (2009) are not in agreement with our photoionization model. Such an explanation, obviously, does not hold for oxygen, because both \Op\ and \Opp\ lines are measured. Yet, there is a 0.1 dex difference between the model value and the value derived by Bresolin (2007). The reason is that this model is not perfect, because it cannot reproduce at the same time all the line ratios involving  \Oiii.
 
\begin{figure}[!h]
\centering
\includegraphics[scale=0.5]{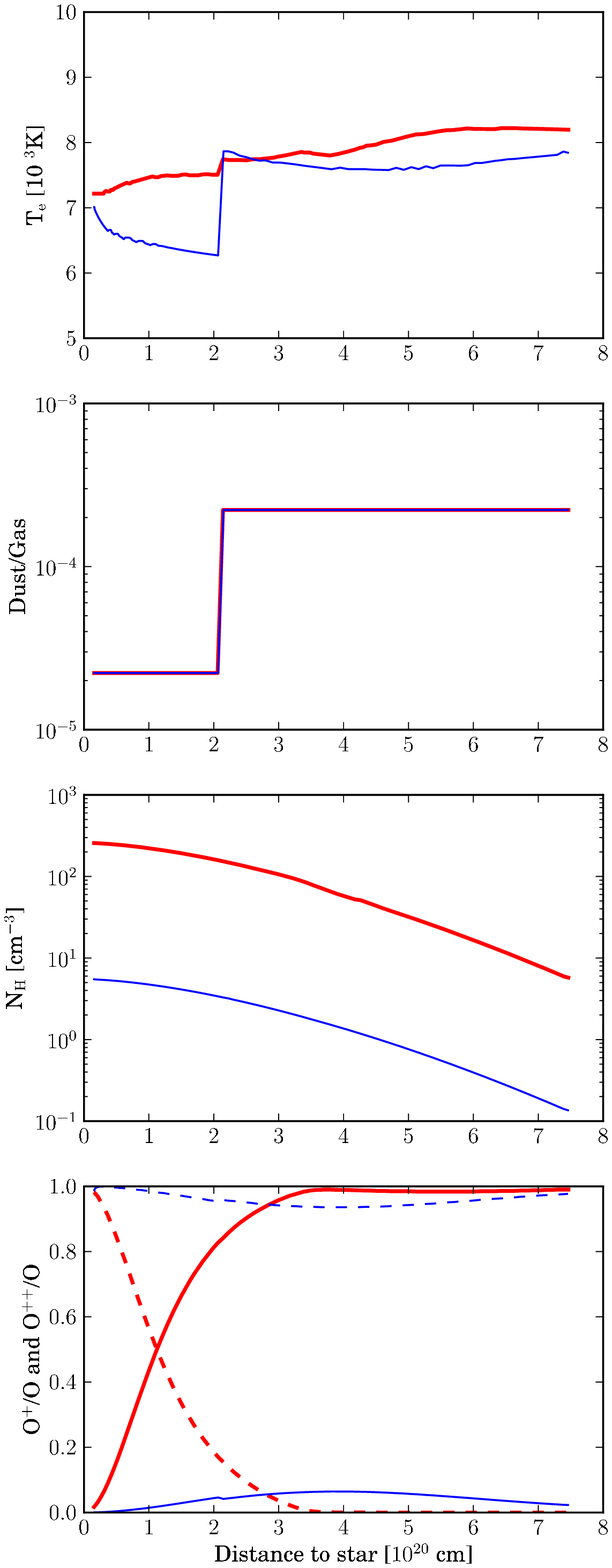}
\caption{Radial distribution of the electron temperature, the dust-to-gas ratio,  the hydrogen density, and the relative abundances of \Op\ (dashed curve) and \Opp\ (continuous curve). The thick red curves correspond to Model M1 (or the  high-density component of model M2) while the thin blue curves correspond to the low density component of model 2.
\label{fig:radialvar}}
\end{figure}

In Fig. \ref{fig:radialvar} we show the radial distribution of the electron temperature, the dust-to-gas ratio,  the hydrogen density, and the relative abundances of \Op\ and \Opp\ in model M1 (red curves).  We see that the radial gradient of the electron temperature is mild, with a temperature of $\simeq$ 7200\,K in the central zone and $\simeq$ 7700\,K in the outer zone. This is what we would have expected for an \hii\ region of roughly solar metallicity. The important point is that this temperature distribution is fully compatible with the line intensities observed both in the optical and in the infrared. The small increase of the electron temperature at $2\times 10^{20}$ cm is due to photoelectric heating of the gas by dust grains. We have checked that the amount of dust present in the model is compatible with the extinction $C(\Hb)\simeq 0.3$ estimated for this object from the Balmer lines.

We note that the intensity of the \Oi\ line is smaller by a factor  $\sim$10 compared to the observations. This does not represent a real worry,  since \oi\  is essentially produced in the ionization front (i.e. not in the ionized regions as all the remaining lines), and we did not attempt to model this zone in detail.

The discussion of the recombination lines is postponed to Sect. \ref{sec:recomb}.

\subsection{A composite model}
\label{sec:composite}

One problem that still needs examination is the density structure. Our best model, M1,  requires a volume filling factor of 0.005. This is not uncommon for giant \hii\ regions (eg. Luridiana et al 1999; Stasi\'nska \& Schaerer 1999). Of course, using a filling factor is an approximation. There has to be some diffuse gas between the filaments. We have to check whether such gas affects the intensities of the observed emission lines. We have thus computed additional low-density models with the same abundances at the ``high'' density, main model. The computations are performed until the outer radius of the nebula is reached (so, for low densities, these additional models are density bounded). A composite model is then obtained by adding the fluxes of the main, ``high'' density model, with one of those additional models. This is, of course, a simplification of a more realistic model where the density distribution would be filamentary (as in the model of Jamet et al. 2006), but it is sufficient for our purposes. The results of our ``best fit'' composite model, M2, are shown in Fig. \ref{fig:compar2} and Table \ref{tab:spectrmodobs}. 
The high- and low-density components have  $A_1=235$ and $A_1=5$ cm$^{-3}$, and filling factors of 0.003 and 0.5,   respectively.

Characteristics of the corresponding low-density model are  plotted in Fig. \ref{fig:radialvar}, with blue curves. From  Fig. \ref{fig:compar2} we see that  our best composite model fits the data even better than model M1. In particular, the \nioni{S}{iv}10.5 / \nioni{S}{iii}18.7  and the \Oiii/\Hb\ ratio from Bresolin (2007) are now well fitted. This is not too surprising, since we have allowed more parameters to vary and the resulting model should be closer to reality. Since, for the composite model, we have adopted the same dust-to-gas ratio distribution as for model M1, the jump in electron temperature at the place where grains appear is much higher than in model M1, because, as a consequence of the higher local ionization parameter due to the lower gas density, the effect of heating by grains is enhanced (Stasi\'nska \& Szczerba 2001). Note that, in the central zone, devoid of grains, the temperature of the low-density model is lower than in the case of model M1. This is because oxygen is entirely in the form of \Opp, the most efficient coolant.
In spite of the significant differences in temperatures with model M1, the overall ``temperature fluctuation'' $t^2$ (as defined by Peimbert 1967 and computed by us integrating the temperature weighted by the product $n_e n_H$ over the emitting volume) is very small in this composite medel (less than 0.001).
We have to note that, although other composite models can probably be found to fit all the observations equally well, they will not differ significantly from M2 because of the very strong constraints imposed on each zone of the nebula by the observations.

The chemical composition of the composite model is : H$=12$,  He$=11.0$,  C$=8.66$,  N$=7.74$, O$=8.57$,  Ne$=7.84$,  S$=6.87$,   Cl$=5.27$,  Ar$=6.23$, Fe$=5.87$.

We do not attempt here to assign error bars to the abundances  derived from the model, as this would  require to perfectly fit the models to various combinations of the extreme values of intensities allowed by the observations. This represents a very time consuming task, because line intensities are not directly proportional to abundances.  Fortunately, we do not need to go through such a lengthy procedure. What we wanted is to see whether we can obtain at least one photoionization model including all the classical ingredients which would fit \textit{all} the observations, \textit{including} the new infrared data, which put strong constraints on the temperature. The fact that we succeeded argues against the presence of important temperature inhomogeneities in this object.


\subsection{Recombination lines and temperature fluctuation issues}
\label{sec:recomb}

Let us now look at the recombination lines of C and O as predicted by our models. 
The \ioni{C}{ii} $\lambda$ 4267  line is well reproduced by our models, which indicates that the ionization correction factors  used by both Bresolin (2007) and Esteban et al. (2009) to derive the carbon abundance  are in agreement with our model predictions. This, however, does not mean that the carbon abundance is correct, since one might expect a similar problem with the carbon recombination line as with the oxygen ones. Indeed, the sum of the intensities of multiplet 1 of  \ioni{O}{ii}  (labelled \ioni{O}{ii}  $\lambda$ 4651 in Table \ref{tab:spectrmodobs}) is about 4 times smaller than the observed value (if we use Eq. 11 from Peimbert et al. 2005 to convert the intensity of \ioni{O}{ii}  $\lambda$ 4649 observed by Esteban et al. 2009 into the sum of the multiplet). This is because the oxygen abundance in the models is fixed by fitting the observed intensities of the forbidden lines. Thus, the oxygen abundance discrepancy between the recombination lines and collisionally excited lines appears to be of a factor of 0.6 dex\footnote{This is larger than the value of 0.36 dex derived by Esteban et al. (2009) from their observations due to the conjunction of several causes such as the fact that our model attempts to fit the observation of Bresolin (2007) rather than those of Esteban et al. (2009), and that Esteban (private communication) did not employ the same procedure to estimate the sum of the multiplet.}. That the models underpredict the value for \ioni{O}{ii}  $\lambda$ 4651 is not surprising, since they are chemically homogeneous and do not experience sufficient temperature variations to explain the \textit{observed} \Opp/\Hp\ abundance discrepancy. 

We have made some experiments by enhancing the  oxygen abundance so as to reproduce   \ioni{O}{ii}  $\lambda$ 4651. Oxygen cooling is then so important that it lowers the gas temperature to such a point where the \Oiii\ line becomes severely \textit{underestimated}. In addition, if the oxygen abundance is enhanced by a factor of 3 of 4 with respect to the solar value, from what is known of stellar nucleosynthesis and galactic chemical evolution, this should also be the case of other $\alpha$-elements, such as neon, sulfur and argon. In this case, the intensitiy of the  \nioni{S}{III} $\lambda$18.7$\mu$m line becomes largely overestimated.

Our models also do not explain the Balmer jump temperature of  $5000 \pm 800$ K obtained by Bresolin (2007), which led to his  $t^2=0.06$. However, it must be noted that the spectral resolution of Bresolin's observations was not sufficient to reach an accurate value of the level of the continuum close to the Balmer discontinuity as well as of the intensity of the high-order Balmer lines, so that the Balmer jump with respect to adjacent hydrogen lines could be easily overestimated by 20\%, leading in that case to a Balmer jump temperature close to 7000 K. The spectrum of Esteban et al. (2009) is an echelle spectrum with much higher spectral resolution but poor signal-to-noise ratio in the continuum, so unfortunately the Balmer jump could not be measured there.

\section{Conclusion}
\label{sec:conclusion}

In order to get more insight into the ``temperature fluctuation'' and the ``abundance discrepancy'' problems we have performed a detailed study of the giant \hii\ region H\,1013, for which excellent spectroscopic data exist, and which has been previously diagnosed for the presence of temperature fluctuations. We have first completed the observational data base with our own mid-infrared measurements using the \textit{Spitzer Space Telescope}, which give us important additional constraints, especially on the temperature structure. 

We then made a first estimate of the physical parameters (electron density and temperature) from different line ratios including the ones involving  the infrared data using the PyNeb software (Luridiana et al. 2012). This indicated that temperature fluctuations, even if present, would be much smaller than $t^2$ = 0.06 invoked by Bresolin (2007) from the Balmer jump temperature or $t^2$ = 0.04 from  the \Opp\ abundance discrepancy obtained by comparison of the recombination and collisionally excited lines from the high resolution spectrum of Esteban et al. (2009). 

For a deeper insight, one needs to carry out a detailed photoionization analysis in which all the processes that affect the line intensities as well as the observing apertures are taken into account. We performed such a study, using Starburst99 and the Cloudy\_3D package, trying to find  a photoionization model of  H\,1013  that is able to reproduce \textit{all} the observational data (surface brightness distribution, line ratios and stellar features). 

We were able to produce an almost satisfactory model, assuming a volume filling factor of 0.005. Since the use of a filling factor is just  a zero-order approach, we checked the effect of adding a low-density medium to the modelled clumps or filaments. This resulted in an even better fit to the observations.
Of course, as in any study which requires the inclusion of a volume filling factor or -- more realistically -- with a large volume of diffuse gas bathing the filaments,  we are left with the question of what continuously generates those filaments, since their lifetime must be short, as they are not in pressure equilibrium with the surrounding diffuse matter.

The fact that we are able to produce photoionization models that perfectly fit most observational constraints -- which are strong and numerous  -- argues against the presence of significant temperature fluctuations in H\,1013, contrary to what has been advocated previously for this object. The oxygen abundance of our best model  is 12 + log O/H = 8.57, as opposed to the values of 8.73 and 8.93 advocated by Esteban et al. (2009) and Bresolin (2007), respectively.

However, our model is not able to reproduce the intensities of the  oxygen recombination  lines, which is one of the big things that remain to be explained. Our model also does not reproduce the very low Balmer jump temperature inferred by Bresolin (2007) from his data, but we have argued that his value might be in error, possibly due to insufficient spectral resolution of the observations for such a purpose. A new observational study, focussing on the Balmer jump temperature in this nebula, would obviously be very important for our case.

This paper can by no means be considered a solution to the nagging  ``temperature fluctuation'' and  ``abundance discrepancy'' problems, since it studies just one object and does not even provide a complete understanding of the observational data. But it serves as a  warning against too general statements about the physical
conditions in \hii\ regions and too hasty conclusions drawn from their spectra. A promising solution is of course the deviation from a Maxwellian distribution of electron energies, proposed by Nicholls et al. (2012). Work in this direction is needed to show quantitatively that such circumstances i) can occur in \hii\ regions and ii) are able to produce photoionization models  that reproduce a wide range of observational constraints, as we have attempted.

\begin{acknowledgements}
We wish to thank B. Cedr\'es and J. Cepa for sharing with us the ALFOSC images of M101  and Ryszard Szczerba for helping with the \textit{Spitzer} spectrum. 
SS-D acknowledges financial support by  the Spanish Ministerio de Ciencia e Innovaci\'on under
the project AYA2008-06166-C03-01 and the Spanish MICINN under the Consolider-Ingenio 2010 Program 
grant CSD2006-00070: First Science with the GTC  (http://www.iac.es/consolider-ingenio-gtc).
GS and CM acknowledge support from the following Mexican projects:  CB2010:153985, PAPIIT-IN105511
and PAPIIT-IN112911.
This work is based in part on observations made with the Spitzer Space Telescope, which is operated by the Jet Propulsion Laboratory, California Institute of Technology, under a contract with NASA. Support for this work was provided by NASA through an	award issued by	JPL/Caltech.

\end{acknowledgements}


%





%
%

\begin{thebibliography}{}


\bibitem[\protect\citeauthoryear{Aller 
\& Menzel}{1945}]{1945ApJ...102..239A} Aller L.~H., Menzel D.~H., 1945, ApJ, 102, 239 

\bibitem[\protect\citeauthoryear{Baade 
\& Minkowski}{1937}]{1937ApJ....86..119B} Baade W., Minkowski R., 1937, ApJ, 86, 119




\bibitem[\protect\citeauthoryear{Bresolin}{2007}]{2007ApJ...656..186B} 
Bresolin F., 2007, ApJ, 656, 186 


\bibitem[\protect\citeauthoryear{Cedr{\'e}s 
\& Cepa}{2002}]{2002A&A...391..809C} Cedr{\'e}s B., Cepa J., 2002, A\&A, 391, 809 

\bibitem[\protect\citeauthoryear{Cervi{\~n}o et 
al.}{2003}]{2003A&A...407..177C} Cervi{\~n}o M., Luridiana V., P{\'e}rez E., V{\'{\i}}lchez J.~M., Valls-Gabaud D., 2003, A\&A, 407, 177 


\bibitem[\protect\citeauthoryear{Esteban et 
al.}{2009}]{2009ApJ...700..654E} Esteban C., Bresolin F., Peimbert M., 
Garc{\'{\i}}a-Rojas J., Peimbert A., Mesa-Delgado A., 2009, ApJ, 700, 654 


\bibitem[\protect\citeauthoryear{Ferland et 
al.}{1998}]{1998PASP..110..761F} Ferland G.~J., Korista K.~T., Verner 
D.~A., Ferguson J.~W., Kingdon J.~B., Verner E.~M., 1998, PASP, 110, 761 


\bibitem[\protect\citeauthoryear{Garc{\'{\i}}a-Rojas 
\& Esteban}{2007}]{2007ApJ...670..457G} Garc{\'{\i}}a-Rojas J., Esteban C., 2007, ApJ, 670, 457 

\bibitem[\protect\citeauthoryear{Jamet et 
al.}{2004}]{2004A&A...426..399J} Jamet L., P{\'e}rez E., Cervi{\~n}o M., Stasi{\'n}ska G., Gonz{\'a}lez Delgado R.~M., V{\'{\i}}lchez J.~M., 2004, A\&A, 426, 399 

\bibitem[\protect\citeauthoryear{Jamet et 
al.}{2005}]{2005A&A...444..723J} Jamet L., Stasi{\'n}ska G., P{\'e}rez E., Gonz{\'a}lez Delgado R.~M., V{\'{\i}}lchez J.~M., 2005, A\&A, 444, 723 


\bibitem[\protect\citeauthoryear{Kroupa}{2001}]{2001MNRAS.322..231K} Kroupa 
P., 2001, MNRAS, 322, 231 




\bibitem[\protect\citeauthoryear{Luridiana, Peimbert, 
\& Leitherer}{1999}]{1999ApJ...527..110L} Luridiana V., Peimbert M., Leitherer C., 1999, ApJ, 527, 110 

\bibitem[]{} Luridiana,V.,   Morisset, C.,   Shaw, R. A. 2012, in IAU Symp. 283 "Planetary Nebulae: an Eye to the Future" 


\bibitem[\protect\citeauthoryear{Mesa-Delgado et 
al.}{2012}]{2012MNRAS.426..614M} Mesa-Delgado A., N{\'u}{\~n}ez-D{\'{\i}}az 
M., Esteban C., Garc{\'{\i}}a-Rojas J., Flores-Fajardo N., 
L{\'o}pez-Mart{\'{\i}}n L., Tsamis Y.~G., Henney W.~J., 2012, MNRAS, 426, 
614 

\bibitem[\protect\citeauthoryear{Morisset}{2006}]{2006IAUS..234..467M} 
Morisset C., 2006, IAUS, 234, 467 

\bibitem[\protect\citeauthoryear{Morisset 
\& Georgiev}{2009}]{2009A&A...507.1517M} Morisset C., Georgiev L., 2009, A\&A, 507, 1517 

\bibitem[\protect\citeauthoryear{Nicholls, Dopita, 
\& Sutherland}{2012}]{2012ApJ...752..148N} Nicholls D.~C., Dopita M.~A., Sutherland R.~S., 2012, ApJ, 752, 148 

\bibitem[\protect\citeauthoryear{Osterbrock}{1989}]{1989agna.book.....O} 
Osterbrock D.~E., 1989, agna.book

\bibitem[\protect\citeauthoryear{Peimbert 
\& Costero}{1969}]{1969BOTT....5....3P} Peimbert M., Costero R., 1969, BOTT, 5, 3 


\bibitem[\protect\citeauthoryear{Peimbert}{1967}]{1967ApJ...150..825P} 
Peimbert M., 1967, ApJ, 150, 825 

\bibitem[\protect\citeauthoryear{Peimbert 
\& Peimbert}{2011}]{2011RMxAC..39....1P} Peimbert M., Peimbert A., 2011, RMxAC, 39, 1 











\bibitem[\protect\citeauthoryear{Stasi{\'n}ska}{2005}]{2005A&A...434..507S} Stasi{\'n}ska G., 2005, A\&A, 434, 507 


\bibitem[\protect\citeauthoryear{Stasi{\'n}ska}{2004}]{Sta2004} Stasi\'nska, G. 2004, in Cosmochemistry: The Melting Pot of the
Elements, ed. C. Esteban, R. J. Garc\'ia L\'opez, A. Herrero, \&
F. S\'anchez (Cambridge: Cambridge Univ. Press), 115

\bibitem[\protect\citeauthoryear{Stasi{\'n}ska}{2009}]{2009elu..book....1S} 
Stasi{\'n}ska G., 2009, elu..book, 1 

\bibitem[\protect\citeauthoryear{Stasi{\'n}ska 
\& Schaerer}{1999}]{1999A&A...351...72S} Stasi{\'n}ska G., Schaerer D., 1999, A\&A, 351, 72

\bibitem[\protect\citeauthoryear{Stasi{\'n}ska 
\& Szczerba}{2001}]{2001A&A...379.1024S} Stasi{\'n}ska G., Szczerba R., 2001, A\&A, 379, 1024 


\bibitem[\protect\citeauthoryear{Stasi{\'n}ska et 
al.}{2007}]{2007A&A...471..193S} Stasi{\'n}ska G., Tenorio-Tagle G., Rodr{\'{\i}}guez M., Henney W.~J., 2007, A\&A, 471, 193 

\bibitem[\protect\citeauthoryear{Stasi{\'n}ska et 
al.}{2010}]{2010A&A...511A..44S} Stasi{\'n}ska G., et al., 2010, A\&A, 511, A44 

\bibitem[\protect\citeauthoryear{Storey 
\& Hummer}{1995}]{1995MNRAS.272...41S} Storey P.~J., Hummer D.~G., 1995, MNRAS, 272, 41 







\bibitem[\protect\citeauthoryear{Torres-Peimbert 
\& Peimbert}{2003}]{2003IAUS..209..363T} Torres-Peimbert S., Peimbert M., 2003, IAUS, 209, 363 

\bibitem[\protect\citeauthoryear{Tsamis 
\& P{\'e}quignot}{2005}]{2005MNRAS.364..687T} Tsamis Y.~G., P{\'e}quignot D., 2005, MNRAS, 364, 687 





%
\end{thebibliography}
\end{document}